\documentclass[%
amsmath,amssymb,aps,showkeys, nofootinbib, twocolumn,superscriptaddress
]{revtex4-2}

\usepackage{graphicx}
\graphicspath{ {./figures/} }
\usepackage{dcolumn,hyperref}
\usepackage{bm}
\begin{document}

\title{Role of non-magnetic spacers in the magnetic interactions of antiferromagnetic topological insulators MnBi$_{4}$Te$_{7}$ and MnBi$_{2}$Te$_{4}$}

\author{Bing Li}
\email[Current address: ]{Neutron Scattering Division, Oak Ridge National Laboratory, Oak Ridge, TN, 37831, USA}
\affiliation{Ames National Laboratory, Ames, IA, 50011, USA}
\affiliation{Department of Physics and Astronomy, Iowa State University, Ames, IA, 50011, USA}

\author{D. M. Pajerowski}
\affiliation{Oak Ridge National Laboratory, Oak Ridge, TN, 37831, USA}

\author{J.-Q.~Yan}
\affiliation{Oak Ridge National Laboratory, Oak Ridge, TN, 37831, USA}

\author{R.~J.~McQueeney}
\affiliation{Ames National Laboratory, Ames, IA, 50011, USA}
\affiliation{Department of Physics and Astronomy, Iowa State University, Ames, IA, 50011, USA}

\date{\today}
\begin{abstract}
MnBi$_{4}$Te$_{7}$ belongs to a family of antiferromagnetic topological insulators.  It forms a natural heterostructure of magnetic (septuple) and non-magnetic (quintuple) topological blocks. 
Here, we explore the magnetism and magnetic interactions in this compound using inelastic neutron scattering. 
We find that the interlayer magnetic coupling is much weaker in MnBi$_{4}$Te$_{7}$ as compared to  MnBi$_{2}$Te$_{4}$ due to the insertion of non-magnetic quintuple layers in the former.
However, other key magnetic energy scales residing within a single septuple block, the single-ion anisotropy and long-range intralayer exchanges, are essentially the same. 
This identifies a transferable set of magnetic interactions applicable to the extended family of magnetic topological insulators based on MnBi$_2$Te$_4$-Bi$_2$Te$_3$ heterostructures.
\end{abstract}

\maketitle
\section{Introduction}
\label{sec:Intro}
The MnBi$_2$Te$_4$ (MBT) compound is the first known intrinsic antiferromagnetic topological insulator~\cite{Otrokov17, Otrokov19a, Yan19b} and has been utilized as a platform to interrogate the novel phenomena in topological phases such as axion and Chern insulators~\cite{Zhang19, Deng20, Liu20, Gao21,Chong24}. The key to accessing these topological properties is the van der Waals (vdW) coupling between structural blocks, resulting in weak interlayer antiferromagnetic interactions. Combined with an uniaxial magnetic anisotropy of the Mn sublattice, MBT provides easily accessible metamagnetic spin-flop and forced ferromagnetic phases that host unique topological phenomena~\cite{Tan20}.  

Much less explored are the extended family of MnBi$_{2+2n}$Te$_{4+3n}$ compounds where MBT septuple blocks are separated by $n$ non-magnetic Bi$_2$Te$_3$ quintuple topological insulator (TI) blocks \cite{Vidal19, Wu19, Yan20, Ding20, Klimovskikh20, Ceccardi23}. 
A single septuple block is ferromagnetic (FM), while the interlayer magnetic coupling is antiferromagnetic (AF), at least for $n=$0, 1, and 2. 
The insertion of non-magnetic spacer blocks increases the Mn sublattice interlayer distance and, therefore, is expected to reduce the interlayer magnetic coupling, affecting the landscape of magnetic phases and magnetization dynamics~\cite{Klimovskikh20, Wu20, Tan20, Ge22, Yu23, Cui23,Guo24}.
The MBT family provides a means for controlled studies of the magnetization dynamics where interlayer couplings can be tuned from weak (for MnBi$_2$Te$_4$) to essentially zero (for MnBi$_8$Te$_{13}$)~\cite{Wu20}.
In the latter $n=3$ system, long-range dipolar interactions dominate the interlayer coupling, resulting in fragile ferromagnetism~\cite{Hu20}.

It has been suggested that large $n$ MBT compounds form a new class of magnets, called single-layer magnets \cite{Wu20}, which consist of a periodic array of magnetic layers with vanishing interlayer coupling. Thus, in addition to the interesting topological properties, the MBT family also allows for the exploration of the interplay between dimensionality and magnetism in these unique van der Waals bonded natural heterostructures.

In the quest to understand the evolution towards single-layer magnets, here we describe inelastic neutron scattering (INS) measurements on the $n=1$ compound MnBi$_4$Te$_7$, and compare these results to previous measurements on the $n=0$ compound MnBi$_2$Te$_4$\cite{Li20, Li21}.
In those measurements, we found MnBi$_2$Te$_4$ to possess long-range and competing intralayer magnetic interactions, giving rise to a linear sawtooth-like spin wave dispersion.
The uniaxial single-ion anisotropy align the Mn spins along the crystallographic $c$-axis, and the relatively strong AF interlayer coupling across the vdW gap ($J_c$) results in the AF stacking of FM layers. The metamagnetic spin-flop transition in a magnetic field applied perpendicular to the layers can be described by a microscopic model using an Hamiltonian with exchange interactions and a single-ion anisotropy, using parameters determined from INS measurements~\cite{Lai21}. The N\'eel temperature for MnBi$_2$Te$_4$, $T_{\text{N}}=24$~K, is close to $T^{\text{MF}}_{\text{N}}=-\frac{S(S+1)}{3 k_{\text{B}}}\sum_i{J_i} \approx 19$~K estimated from the mean-field theory, where the dominant energy scale is the FM nearest intralayer interaction $J_1$.

Despite having the same A-type AF magnetic ground state, however, in MnBi$_4$Te$_7$ the spin-flop transition is absent~\cite{Tan20}. Together with a reduction of $T_{\text{N}} = 14$~K, these suggest modifications of the magnetic interactions due to the presence of the non-magnetic spacers. In our neutron experiments described here, we find that MnBi$_4$Te$_7$ has a strongly suppressed interlayer spin wave bandwidth and a reduction in the spin gap compared to MnBi$_2$Te$_4$. 
Both observations are consistent with vanishing interlayer exchange coupling and a transferable single-ion anisotropy.
Without the influence from the interlayer couplings, the interactions within a single septuple block could be more reliably determined using the linear spin wave theory (LSWT) analysis based on the Heisenberg model. 
This analysis confirms that long-range and competing intralayer interactions found in MnBi$_2$Te$_4$ are intrinsic to the septuple blocks in all MBT family members.

Overall, this suggests that MnBi$_4$Te$_7$ and successive $n>1$ MBT family members are quite close to the 2D limit. Considering that INS is a bulk measurement, we propose that MnBi$_4$Te$_7$ serves as a suitable system to probe 2D magnetism, which is not only interesting theoretically but also relevant for the development of few-layer magnetic TI devices~\cite{Cui23, Lei21, Yang21, Lei22}.

\begin{figure*}[ht]
\centering
\includegraphics[width=0.82\paperwidth]{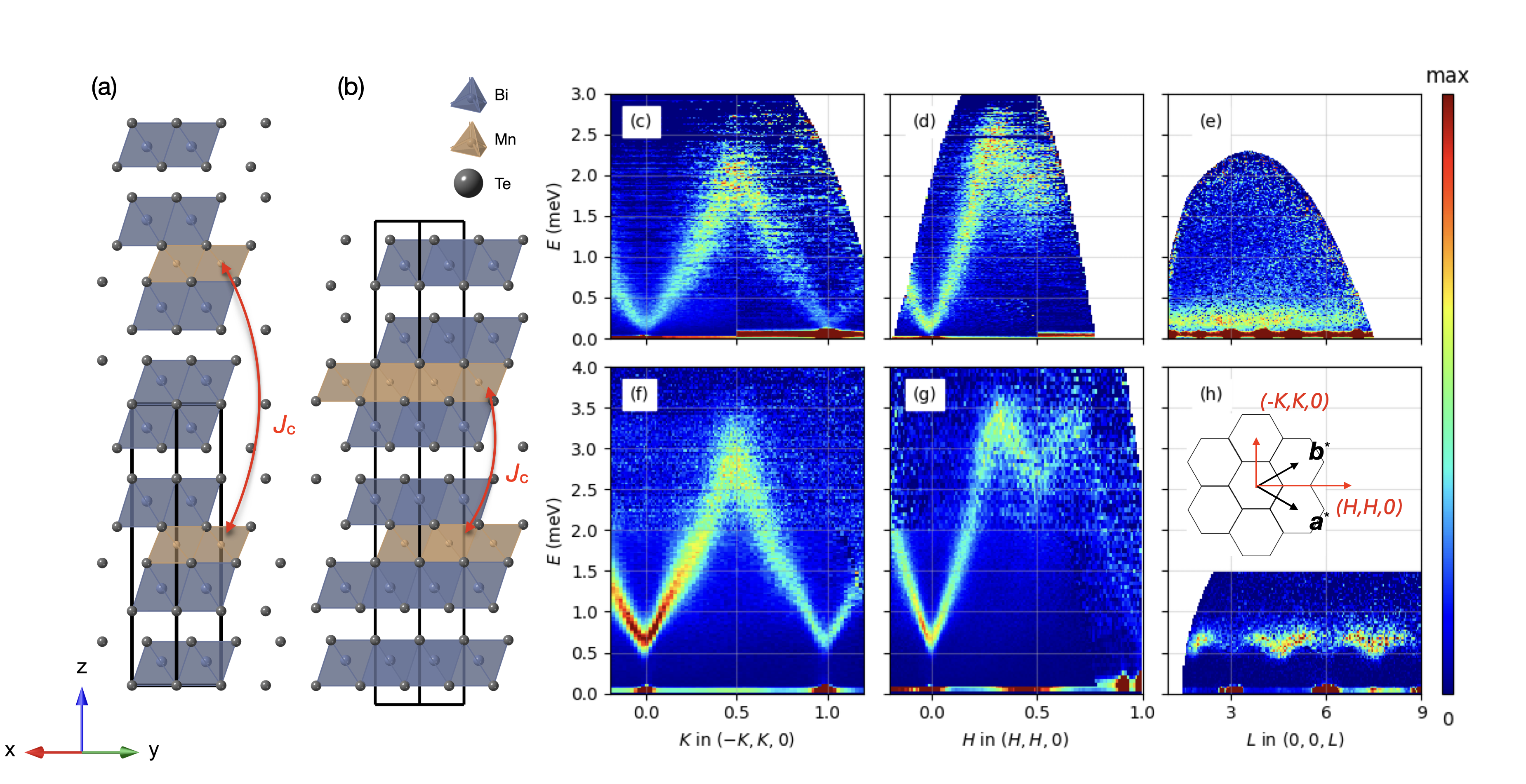}
\caption{Crystal structure of (a) MnBi$_4$Te$_{7}$ and (b) MnBi$_2$Te$_{4}$ with $J_c$ indicating the distance of the nearest interlayer neighbor. (c-e) Spin waves in MnBi$_4$Te$_{7}$ measured at $T=2$~K with $E_{\rm{i}}=3.32$ meV. Measurement with $E_{\rm{i}}=1.55$~meV is overlaid on top in (c) and (d) for better visualization of the spin gap at low $E$. (f-h) Spin waves in MnBi$_2$Te$_{4}$ measured at $T=2$~K with $E_{\rm{i}}=6.6$~meV, with $E_{\rm{i}}=3.3$~meV measurement overlaid on top in (f) and (g). The intensity in (c-h) is multiplied by energy transfer $E$ to enhance the contrast at higher $E$. Inset of (h) shows the 2D projection of the Brillouin zone and the high-symmetry directions.}
\label{fig:disp}
\end{figure*}

\section{Experimental Details}
\label{sec:exp}

Single-crystal samples of MnBi$_4$Te$_7$ and MnBi$_2$Te$_4$ were grown out of a Bi-Te flux~\cite{Yan20}. The elemental analysis and X-ray and neutron diffraction results reveal that the Mn site hosts extra Bi and Mn vacancies, giving a chemical formula of Mn$_{0.8}$Bi$_{4.2}$Te$_7$.
MnBi$_2$Te$_4$ crystallizes in space group $R\bar3m$ (\#$166$), with lattice parameters $a=b=4.33$~\AA, $c=40.91$~\AA~whereas MnBi$_4$Te$_7$ crystallizes in space group $P\bar3m1$ (\#$164$), with lattice parameters $a=b=4.366$~\AA, $c=23.80$~\AA. 
The essential difference between the two chemical structures is the close-packed stacking of magnetic Mn sublattice in the former and the simple hexagonal stacking in the latter. A comparison of the two crystal structures is shown in Fig.~\ref{fig:disp}(a) and (b).

For both compounds, the 2D projection of the Brillouin zone is hexagonal and we define reciprocal space vectors $\bm{q}=\frac{4\pi}{\sqrt{3} a}(H\hat{\bm{a}}^* + K\hat{\bm{b}}^*) + \frac{2\pi}{c}L\hat{\bm{c}}^*$ and special points $\it{\Gamma}$ $=(0,0,0)$, $M=(1/2,0,0)$, $K=(1/3,1/3,0)$.
SQUID measurements for MnBi$_4$Te$_7$ confirm that AFM ordering occurs at $T_{\rm{N}}=13.7$~K and neutron diffraction finds a magnetic propagation vector of $\bm{\tau}=(0,0,\frac{1}{2})$,  consistent with the A-type AFM order~\cite{Yan20}. 
Previous reports of MnBi$_2$Te$_4$ find $T_{\rm{N}}=24.7$~K and $\bm{\tau}=(0,0,\frac{3}{2})$ for the A-type order of Mn layers with rhombohedral stacking~\cite{Yan19b}.

Similar to MnBi$_2$Te$_4$, the INS measurements on MnBi$_4$Te$_7$ were performed using the Cold Neutron Chopper Spectrometer (CNCS) at the Spallation Neutron Source in Oak Ridge National Laboratory. Co-aligned single-crystal samples with a total mass of 503.7~mg~\cite{SM} were mounted to a top-loading orange cryostat. Measurements were performed using the incident energies $E_{\rm{i}}=3.32$ and $1.55$~meV at base temperatures $T=2$~K. Measurements at higher temperatures were also taken for the estimation of background signals.

\section{Results}
\label{sec:baseT}
Spin waves along high-symmetry directions in MnBi$_4$Te$_7$ are shown in Fig.~\ref{fig:disp}(c-e), compared to spin waves in MnBi$_2$Te$_4$ shown in Fig.~\ref{fig:disp}(f-h) which are originally reported in Ref.~\cite{Li21}. 
The intensity is multiplied by energy transfer $E$ to enhance the contrast at higher $E$. 
Square brackets are used to denote the range of summation in $\bm{q}$ in reciprocal lattice units. Three orthogonal axes used to denote $\bm{q}$ are $(H,H,0)$, $(-K,K,0)$, and $(0,0,L)$.
For MnBi$_4$Te$_7$ in Fig.~\ref{fig:disp}(c) and (d), $L=[2, 6]$.
$H = [-0.05, 0.05]$ in (c), and $K = [-0.05, 0.05]$ in (d). In (e), $H=K=[-0.04, 0.04]$.
For MnBi$_2$Te$_4$ in Fig.~\ref{fig:disp}(f) and (g), $L=[0, 20]$ for $E_{\rm{i}}=6.6$~meV data and $L=[3,12]$ for $E_{\rm{i}}=3.3$~meV data. $H = [-0.05, 0.05]$ in (f), and $K = [-0.05, 0.05]$ in (g). In (h), $H=K=[-0.02, 0.02]$.

Comparison of Figs.~\ref{fig:disp}(c) with \ref{fig:disp}(f) and Figs.~\ref{fig:disp}(d) with \ref{fig:disp}(g) indicate that the in-plane dispersions of MnBi$_4$Te$_{7}$ and MnBi$_2$Te$_{4}$ are quite similar.
One obvious difference is the slightly lowered spin wave energy maximum (from $3.3$~meV to $2.3$~meV), and concomitantly the smaller spin gap in MnBi$_4$Te$_{7}$. A comparison of the interlayer dispersions along $L$ in Figs.~\ref{fig:disp}(e) with \ref{fig:disp}(h) highlights that the interlayer coupling is strongly reduced in MnBi$_4$Te$_{7}$.  

We discuss these features of the spin wave spectrum using LSWT based on the following Heisenberg model,
\begin{equation}
    \mathcal{H} = \sum_{\langle ij \rangle,\parallel} J_{ij} \bm{S}_{i} \cdot \bm{S}_{j} - D \sum_{i}(S^z_i)^2.
\label{eqn:heisenberg_147}
\end{equation}
This model is similar to that used in the analysis of MnBi$_2$Te$_4$ \cite{Li21}. The difference is that we use a 2D model for MnBi$_4$Te$_7$ where the interlayer coupling terms $J_{\rm{c}} \sum_{\langle ij \rangle,\perp}\bm{S}_{i} \cdot \bm{S}_{j}$ are not included. In the next section, we justify this choice by showing that the reduced dispersion along $L$ is too small to resolve $J_{\rm{c}}$ with our current data.

\subsection{Interlayer couplings}
\label{sec:Jc}

\begin{figure*}
\centering
\includegraphics[width=0.82\paperwidth]{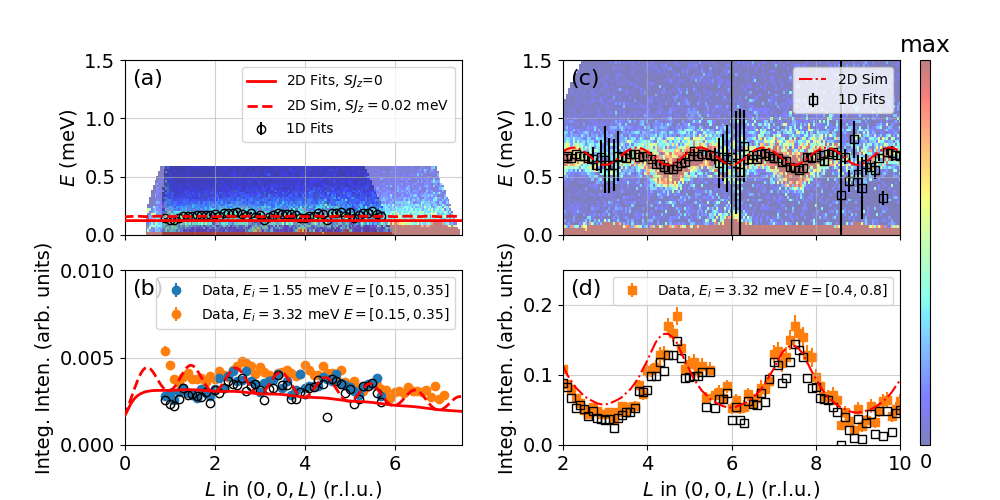}
\caption{Integrated INS intensities and spin wave dispersion along $(0,0,L)$~direction for MnBi$_4$Te$_7$ (a) and MnBi$_2$Te$_4$ (c), similarly to data plotted in Fig.~\ref{fig:disp}(e) and (f). Data with $E_{\rm{i}}=1.55$~meV are overplotted on data with $E_{\rm{i}}=3.32$~meV in (a). (b) and (d) show intensity variation when summed over the energy range $E=[0.15,0.35]$~meV and $E=[0.4,0.8]$~meV for MnBi$_4$Te$_7$ and MnBi$_2$Te$_4$, respectively. Black open symbols with error bars are peak centers in (a) and (c), and integrated intensity in (b) and (d), obtained from fitting individual constant-$q$ line cuts. (circles for MnBi$_4$Te$_7$ and squares for MnBi$_2$Te$_4$.) The large error bars in (c) result from the vanishing intensities when approaching $L=3n$. 
Red curves show the simulated dispersion and intensity variation from fitting to the 2D slice of data. Notice that 1D and 2D fitting give consistent results in MnBi$_2$Te$_4$. Therefore, the intensity variation is used for an estimation of the upper limit of the interlayer interaction $J_{\rm{c}}$ in MnBi$_4$Te$_7$ as shown in (b).}
\label{fig:dispL}
\end{figure*}

In Fig.~\ref{fig:dispL}, the experimental data of the interlayer dispersion is compared to simulations of the data from linear spin wave theory. For MnBi$_2$Te$_4$, one finds that the AFM interlayer coupling has two effects. It creates an oscillatory dispersion along $(0,0,L)$ [Fig.~\ref{fig:dispL}(c)] as well as a variation of scattering intensities [Fig.~\ref{fig:dispL}(d)].  In the latter case, the intensities having maxima at $L=3n\pm\frac{3}{2}$ and minima at $L=3n$ which result from the AFM propagation vector of $\bm{\tau}=(0,0,3/2)$, corresponding to staggered magnetization of hexagonal close-packed magnetic layers. 

In MnBi$_4$Te$_7$, however, neither of the two effects are clearly observed. Analysis on the spin wave dispersion along $(0,0,L)$ gives a single-ion anisotropy $SD_z=0.077(1)$~meV, similar to the value $SD_z=0.0814$~meV in MnBi$_2$Te$_4$ reported in Ref.~\cite{Li21}, and an unstable interlayer interaction $SJ_{\rm{c}}=0.001(1)$~meV~\cite{SM}, in comparison to $SJ_{\rm{c}}^{124}=0.065(2)$~meV~\cite{Li21}. Integrated intensities at low $E=[0.15, 0.35]$~meV are plotted in Fig.~\ref{fig:dispL}(b) for MnBi$_4$Te$_7$ along $(0,0,L)$ together with the simulated intensities with $SJ_{\rm{c}}=0$ and $SJ_{\rm{c}}=0.02$~meV. 
Due to the simple stacking of hexagonal magnetic layers in MnBi$_4$Te$_7$, the intensity maxima are expected at $L=n \pm \frac{1}{2}$. 
However, no distinctive intensity variations are observed with the current counting statistics. 
For example, with $SJ_{\rm{c}}\geq0.02$~meV, Fig~\ref{fig:dispL}(b) shows that the variation of intensities should be significant enough to be observed. Therefore we estimate an upper limit of $SJ_{\rm{c}}=0.02$~meV for the interlayer coupling in MnBi$_4$Te$_7$.  

It is worth mentioning that the difference in magnetic layer stacking results in each Mn ion in MnBi$_4$Te$_7$ having only two interlayer nearest-neighbors, whereas in MnBi$_2$Te$_4$ each Mn has six. 
Thus, the upper limit of the energy scale of the total interlayer coupling in MnBi$_4$Te$_7$ is at least an order of magnitude smaller than MnBi$_2$Te$_4$.
\begin{equation}
    \frac{2 J_{\rm{c}}^{147}}{6 J_{\rm{c}}^{124}}=\frac{2 \times 0.02}{6 \times 0.065} \lesssim 0.10.
\end{equation}

\subsection{Spin gap}
\label{sec:gap}

\begin{figure}[ht]
\centering
\includegraphics[width=0.4\paperwidth]{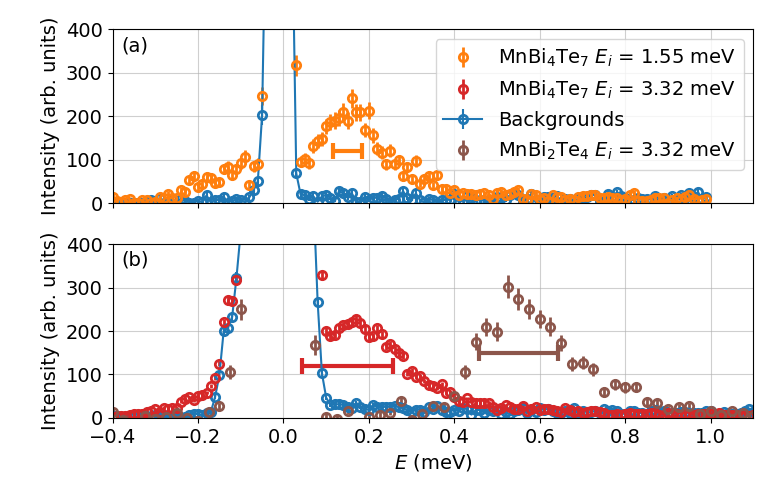}
\caption{Spin gap in MnBi$_4$Te$_7$ at the magnetic $\Gamma$ point, compared to MnBi$_2$Te$_4$. 
Orange and red circles are INS intensities at $\bm{q}=(0,0,2.5)$, having $E_{\rm{i}}=1.55$~meV and $3.32$~meV, respectively.
Blue circles are at $\bm{q}=(0.3,0,2.5)$, demonstrating the  incoherent background. Brown circles are INS intensities at $\bm{q}=(0,0,4.5)$ for MnBi$_2$Te$_4$. The horizontal bars show the instrumental resolution FWHM at peak center positions.}
\label{fig:gap}
\end{figure}

To investigate the single-ion anisotropy $D$ in MnBi$_4$Te$_7$, cuts were made as functions of energy transfer $E$ for data sets with different incident neutron energies. 
Figs.~\ref{fig:gap}(a) and (b) show INS intensities at the bottom of the magnon band (magnetic $\Gamma$ point) $\bm{q}=(0,0,2.5)$~(r.l.u.), with the summation range $H=[-0.04,0.04]$, $K=[-0.03,0.03]$ and $L=[2.3,2.7]$.
The peak at $\Delta \approx 0.2$~meV corresponds to the spin gap. 
Intensities at $\bm{q}=(0.3,0,2.5)$ with the summation range of $H$ offset to $[0.26, 0.34]$ are used to estimate  background signals from incoherent scattering. Whereas the data with $E_{\rm{i}}=3.32$~meV has a coarser resolution than data with $E_{\rm{i}}=1.55$~meV (orange and red), both data sets give a consistent estimate for the spin gap energy. This can be compared to the $\Delta \approx 0.53$ meV for MnBi$_2$Te$_4$ as shown in Fig.~\ref{fig:gap}(b) (brown).

However, in contrast to MnBi$_2$Te$_4$, there is no clean gap in MnBi$_4$Te$_7$ below which magnetic scattering intensities are completely absent. 
The finite spectral weight that is observed down to $E\approx0$ is not only a consequence of the instrumental resolution, considering that the resolution linewidth at $E=0$ is about $0.04$~meV when $E_{\rm{i}}=1.55$~meV. 
Rather, the broadening is caused by the intrinsic damping of the magnons that can be closely approximated as under-damped harmonic oscillators, using the spectral function,

\begin{equation}
\label{eqn:ho}
F(E) \propto \frac{E\cdot \Gamma}{(E^2-E_{0}^2)^2+(E\cdot \Gamma)^2},
\end{equation}

where the damping factor $\Gamma$ satisfies $E_{0}^2\geq\Gamma^2/4$. An accurate estimation of the spin gap and the single-ion anisotropy require 2D fitting of the INS spectra with energy-resolution and $\bm{q}$-binning effects taken into consideration as performed in section~\ref{sec:Jab}. At $\bold{q} = (0,0,2.5)$, the fitting results show that the spectral function is close to being critically damped ($E_{0}\approx\Gamma/2$).

\subsection{Intralayer couplings}
\label{sec:Jab}

Given the two-dimensional nature of the spin excitations in MnBi$_4$Te$_7$ with $J_c\approx 0$, we determine the intralayer magnetic couplings and single-ion anisotropy by using the reduced $\chi^2$ analysis to compare the experimental and simulated spin wave intensities along $(-K,K,0)$ and $(H,H,0)$. Starting with Eqn.~(\ref{eqn:heisenberg_147}), the analytical expression for spin wave dispersion is given by
\begin{equation}
\begin{aligned} 
    E(\bm{q}) &= \sqrt{A(\bm{q})^2 - B(\bm{q})^2},\\
    A(\bm{q}) &= S \left(J(\bm{q})+\frac{1}{2}J(\bm{q}+\bm{\tau})+\frac{1}{2}J(\bm{q}-\bm{\tau})-2J(\bm{\tau}) \right) \\
    & +2SD,\\
    B(\bm{q}) &= S \left(J(\bm{q})-\frac{1}{2}J(\bm{q}+\bm{\tau})-\frac{1}{2}J(\bm{q}-\bm{\tau}) \right).
\label{eqn:AqBqEq_147}
\end{aligned}
\end{equation}
Here $J(\bm{q}) = \sum_{j} J_{0j} e^{i \bm{q} \cdot \bm{R}_j}$ is the Fourier transformation of all neighboring magnetic interactions $J_{ij}$ for a given atom at site $i$, and $\bm{\tau} = (0,0,\frac{1}{2})$ is the magnetic propagation vector. 

The spin wave intensity $I(\bm{q},E)$ is proportional to the sum of the transverse components of the dynamic spin-spin correlation function $S^{xx}(\bm{q}, E)$ and $S^{yy}(\bm{q}, E)$, where
\begin{multline}
\label{eqn:S_tr}
S^{xx}(\bm{q}, E) = S^{yy}(\bm{q}, E)\\
= S\frac{A(\bm{q})-B(\bm{q})}{E(\bm{q})} \delta(E-E(\bm{q}))(1-e^{-\frac{E}{k_{\rm{B}}T}})^{-1},
\end{multline}
and the INS spin wave intensity $I(\bm{q}, E)$ is,

\begin{multline}
\label{eqn:Sqw_147}
   I(\bm{q},E)=S |f(\bm{q})|^{2} \frac{A(\bm{q})-B(\bm{q})}{E(\bm{q})}(1+\hat q_z^2) \\
   \frac{E\cdot \Gamma}{(E^2-E_{0}^2)^2+(E\cdot \Gamma)^2}
   (1-e^{-\frac{E}{k_{\rm{B}}T}})^{-1}.
\end{multline}
$f(\bm{q})$ is the magnetic form factor of Mn$^{2+}$ ion, $\hat q_z$ is the z component of the unit vector in $\bm q$ direction and $(1-e^{-\frac{E}{k_{\rm{B}}T}})^{-1}$ is the Bose thermal factor. The $\delta$-function in Eqn.~(\ref{eqn:S_tr})is replaced by the spectral function in Eqn.~(\ref{eqn:ho}).

One of technical challenges for INS experiment on MnBi$_4$Te$_7$ is that, since Bi and Te atoms are both heavy, the atomic percentage of Mn is much lower compared to MnBi$_2$Te$_4$. 
For samples with identical mass, the strength of magnetic signals in MnBi$_4$Te$_7$ will be 4 times weaker than those in MnBi$_2$Te$_4$. 
Therefore a good estimation of the background signal is crucial for the quantitative analysis. Details of the background subtraction can be found in the supplementary information \cite{SM}, including representative energy cuts of $I(\bm{q},E)$ along $(H,H,0)$ and $(-K,K,0)$.

\begin{figure}[ht]
\centering
\includegraphics[width=0.4\paperwidth]{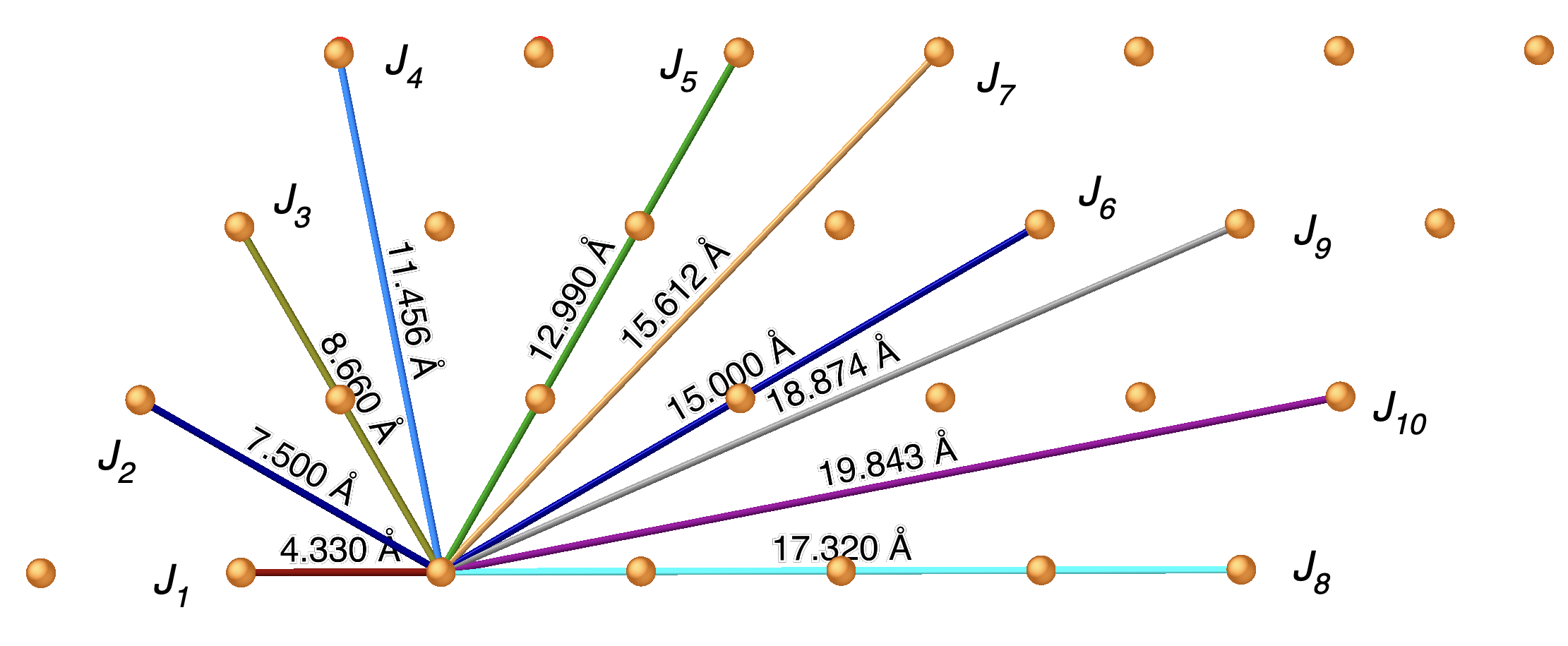}
\caption{Intralayer interactions and their corresponding bond lengths between Mn atoms in MnBi$_2$Te$_4$ and MnBi$_4$Te$_7$.}
\label{fig:nbs}
\end{figure}

To account for magnon damping, cuts of $I(\bm{q}, E)$ were first fit to Eqn.~(\ref{eqn:Sqw_147}) independently to obtain the damping parameter $\Gamma(E)$. $\Gamma(E)$ was then parameterized by the function $\Gamma(E)=c_{1}\arctan(c_{2}E+c_3)+c_4$.
In fits to the full 2D spectra, we keep the value of the single-ion anisotropy $SD_z=0.077$~meV and $J_{\rm{c}}=0$, and let $c_1$, $c_2$, $c_3$, $c_4$ and intralayer interactions $J_{1}, \dots, J_{9}$ vary freely. 
To account for the effects of summation in $\bm{q}$, cuts of the simulations were made with the finite $\bm{q}$-binning identical to the data and compared to the data cuts. A good fit is achieved when the reduced $\chi^2$ (sum of all $\chi^2$ divided by the number of fitting parameters) reaches minimum where reduced $\chi^2 = 10.82$. The intralayer interactions are shown in Fig.~\ref{fig:nbs} and the extracted strength are compiled in Table.~\ref{tab:Jij_147}, together with values in MnBi$_2$Te$_4$ for comparison.

\begin{table*}
    \centering
    \caption{Comparison of the strength of intralayer magnetic coupling $SJ_{ij}$ (meV) in MnBi$_4$Te$_7$ and MnBi$_2$Te$_4$}
    \label{tab:Jij_147}
    \begin{tabular}{  c || c | c | c | c | c | c | c | c | c }
    \hline
     & $SJ_{1}$ & $SJ_{2}$ & $SJ_{3}$ & $SJ_{4}$ & $SJ_{5}$ & $SJ_{6}$ & $SJ_{7}$ & $SJ_{8}$ &$SJ_{9}$ \\
    \hline
     MnBi$_4$Te$_7$   & -0.220(2) & 0.035(2) & -0.012(2) & 0.009(2) & -0.029(1) & 0.014(2) & -0.007(1) & -0.009(1) & -0.009(1) \\
    \hline
    MnBi$_2$Te$_4$    & -0.233(2) & 0.033(2) & -0.007(2) & 0.003(1) & -0.016(2) & -0.013(2) & -0.008(1) & -- & -- \\ 
    \hline
    \end{tabular}
\end{table*}

\section{Discussion}
\label{sec:discussion}
In MnBi$_2$Te$_4$, previous reports find that the interlayer magnetic coupling $J_{\rm{c}}$ is relatively strong considering van der Waals nature of the magnetic coupling between MnBi$_2$Te$_4$ blocks. 
The large spin gap ($\Delta \approx 0.5$ meV) is consistent with the presence of the relatively strong uniaxial anisotropy.
When compared to critical fields for spin-flop and saturation transitions determined from magnetization measurements, we can conclude that the uniaxial anisotropy in MnBi$_2$Te$_4$ has contributions from both single-ion anisotropy and anisotropy of the interlayer exchange coupling.  

In MnBi$_4$Te$_7$, we find the interlayer coupling to be negligible within the sensitivity of our measurement, consistent with the larger separation between magnetic layers. 
However, AF interlayer exchange coupling must still exceed the dipolar interaction that favors a global FM order.
We perform a lattice sum to estimate the net FM interlayer dipolar interaction, $J_{\rm{c}}^{\rm{dip}}= (g\mu_B)^2\frac{\mu_0}{4\pi}\sum_j \frac{r_{0j}^2-3({\bm r}_{0j}\cdot\hat{\bm z})^2}{r_{0j}^5}$. 
By restricting the sum to include only interlayer bonds in the upper-half plane, we obtain $J_{\rm{c}}^{\rm{dip}}\approx -0.003$ meV~\cite{SM}, which is smaller than the estimated upper limit of the total interlayer coupling that includes both exchange and dipolar terms ($J_{\rm{c}}=J_{\rm{c}}^{\rm{ex}}+J_{\rm{c}}^{\rm{dip}} \lesssim 0.01$ meV). 
This suggests that MnBi$_4$Te$_7$ can still host AF order when interlayer AF exchange lies in an estimated range of $ 0.003 \lesssim J_{\rm{c}}^{\rm{ex}} \lesssim 0.013$~meV.  
Ultimately, the interlayer AF exchange is exceeded by the dipolar interactions in MnBi$_8$Te$_{13}$ which crosses over to FM order \cite{Hu20}.

We now discuss the magnetic anisotropy. 
The local atomic configuration surrounding an Mn ion within a single septuple block is essentially the same for MnBi$_2$Te$_4$ and MnBi$_4$Te$_7$.
Thus, we expect that the single-ion anisotropy in the two compounds should be similar and this is confirmed by our experimental data.  
In MnBi$_2$Te$_4$, the overall anisotropy has contributions from interlayer exchange anisotropy~\cite{Li20}, but this contribution will be reduced in MnBi$_4$Te$_7$ due to the overall weaker exchange coupling. 

Magnetization and transport data MnBi$_4$Te$_7$ are consistent with weaker interlayer AF coupling and dominant uniaxial magnetic anisotropy.
Within the Stoner-Wohlfarth model, this weakening results in low-field spin-flip transitions with extremely low coercivity. 
This is borne out by magnetization and Hall measurements \cite{Tan20} where metamagnetism is heavily dominated by uniaxial anisotropy.
 
Finally, our results suggest that the intralayer interactions are transferable across the whole MBT family and play a decisive role in the magnetization dynamics of a single-layer magnet where $n$ is large. The FM nearest-neighbor interaction $J_{1}$ is the dominant interaction, and a competing AFM $J_{2}$ somewhat destabilizes the FM Mn planes. 
Both systems also display a relatively long-range neighbor $J_5$ with comparable strength to $J_{2}$. $J_5$ is in the same crystallographic $a$-axis direction as $J_1$ at distance of $3a$ and provides third order harmonics to $J(q)$ that lead to the distinct linear dispersion relation along $(H,0,0)$.  

\section{Conclusions}
\label{sec:conclusions}
We have measured the spin wave dispersion of the antiferromagnetic topological insulator MnBi$_4$Te$_7$.  
The interactions are strongly two-dimensional in character.  
However, the key magnetic energy scales within a single septuple block, the single-ion anisotropy and the intralayer exchanges, are transferable to all members of the MBT family.

\section*{Acknowledgments}
This work is supported by the U.S. Department of Energy, Office of Basic Energy Sciences, Division of Materials Sciences and Engineering. Ames Laboratory is operated for the U.S. Department of Energy by Iowa State University under Contract No. DE-AC02-07CH11358. This research used resources at the Spallation Neutron Source, a DOE Office of Science User Facility operated by the Oak Ridge National Laboratory. 

\section{Data Availability}
The data that support the findings of this article are openly available \cite{data}.

\bibliographystyle{apsrev4-1}

\end{document}